\documentclass[11pt,a4paper]{article} 
\pdfoutput=1
\usepackage{jheppub,hyperref,mdwlist}
\usepackage{amsmath}
\usepackage{graphicx}
\usepackage{subfigure}
\usepackage{amssymb}
\usepackage{xcolor}
\usepackage{multirow}
\usepackage{cancel}
\usepackage{color}
\usepackage{listings}
\usepackage{xcolor}
\usepackage{float}
\usepackage{slashed}
\usepackage{mathtools}

\usepackage{tikz}
\usetikzlibrary{arrows,shapes}
\usetikzlibrary{trees}
\usetikzlibrary{matrix,arrows} 				
\usetikzlibrary{positioning}				
\usetikzlibrary{calc,through}				
\usetikzlibrary{decorations.pathreplacing}  
\usepackage{pgffor}							

\usetikzlibrary{decorations.pathmorphing}	
\usetikzlibrary{decorations.markings}
\tikzset{
    vector/.style={decorate, decoration={snake}, draw},
	provector/.style={decorate, decoration={snake,amplitude=2.5pt}, draw},
	antivector/.style={decorate, decoration={snake,amplitude=-2.5pt}, draw},
    fermion/.style={draw=black, postaction={decorate},
        decoration={markings,mark=at position .55 with {\arrow[draw=black]{>}}}},
    fermionbar/.style={draw=black, postaction={decorate},
        decoration={markings,mark=at position .55 with {\arrow[draw=black]{<}}}},
    fermionnoarrow/.style={draw=black},
    gluon/.style={decorate, draw=black,
        decoration={coil,amplitude=4pt, segment length=5pt}},
    scalar/.style={dashed,draw=black, postaction={decorate},
        decoration={markings,mark=at position .55 with {\arrow[draw=black]{>}}}},
    scalarbar/.style={dashed,draw=black, postaction={decorate},
        decoration={markings,mark=at position .55 with {\arrow[draw=black]{<}}}},
    scalarnoarrow/.style={dashed,draw=black},
    electron/.style={draw=black, postaction={decorate},
        decoration={markings,mark=at position .55 with {\arrow[draw=black]{>}}}},
	bigvector/.style={decorate, decoration={snake,amplitude=4pt}, draw},
}

\tikzstyle{block} = [draw, rectangle, 
    minimum height=3em, minimum width=6em]

\usepackage{amsmath}
\usepackage{wrapfig}
\usepackage{cleveref}

\newcommand{\be}{\begin{equation}}
\newcommand{\ee}{\end{equation}}
\newcommand{\beq}{\begin{equation}}
\newcommand{\eeq}{\end{equation}}
\newcommand{\bea}{\begin{eqnarray}}
\newcommand{\eea}{\end{eqnarray}}
\newcommand{\besp}{\begin{equation}\begin{split}}
\newcommand{\eesp}{\end{split}\end{equation}}

\newcommand{\Eq}[1]{Eq.~(\ref{#1})}

\newcommand{\Dfbd}{\mathord{\buildrel{\lower3pt\hbox{$\scriptscriptstyle\leftrightarrow$}}\over {D}_{\mu}}}

\def\mO{\mathcal{O}}

\def\Z{\mathbb{Z}}

\def\0{\textbf{0}}
\def\1{\textbf{1}}
\def\2{\textbf{2}}
\def\3{\textbf{3}}
\def\4{\textbf{4}}
\def\5{\textbf{5}}
\def\6{\textbf{6}}
\def\7{\textbf{7}}
\def\8{\textbf{8}}
\def\9{\textbf{9}}

\def\d{\text{d}}

\begin{document}

\title{Primordial black holes from slow phase transitions: a model-building perspective}

\author[a]{Shinya Kanemura,}
\affiliation[a]{Department of Physics, Osaka University, Toyonaka, Osaka 560-0043, Japan}

\author[a, b]{Masanori Tanaka,}
\affiliation[b]{Center for High Energy Physics, Peking University, Beijing 100871, China}

\author[c]{and Ke-Pan Xie}
\affiliation[c]{School of Physics, Beihang University, Beijing 100191, China}

\emailAdd{kanemu@het.phys.sci.osaka-u.ac.jp}
\emailAdd{tanaka@pku.edu.cn}
\emailAdd{kpxie@buaa.edu.cn}

\abstract{
We investigate the formation of primordial black holes (PBHs) through delayed vacuum decay during slow cosmic first-order phase transitions. Two specific models, the polynomial potential and the real singlet extension of the Standard Model, are used as illustrative examples. Our findings reveal that models with zero-temperature scalar potential barriers are conducive to the realization of this mechanism, as the phase transition duration is extended by the U-shaped Euclidean action. We find that the resulting PBH density is highly sensitive to the barrier height, with abundant PBH formation observed for sufficiently high barriers. Notably, the phase transition needs not to be ultra-supercooled (i.e. the parameter $\alpha\gg1$), and the commonly used exponential nucleation approximation $\Gamma(t)\sim e^{\beta t}$ fails to capture the PBH formation dynamics in such models.
}

\begin{flushright}
OU-HET-1226
\end{flushright}

\maketitle
\flushbottom

\section{Introduction}

Due to their profound cosmological and astrophysical implications, primordial black holes (PBHs) have attracted significant research attention~\cite{Carr:2020gox,Green:2020jor,Escriva:2022duf,LISACosmologyWorkingGroup:2023njw}. Those hypothetical black holes originate in the early Universe shortly after the Big Bang, long before the existence of stars and galaxies. Cosmic first-order phase transitions (FOPTs) provide a favorable environment for PBH formation: the Universe transitions from a higher energy false vacuum to a lower energy true vacuum via tunneling, resulting in the true vacuum bubbles nucleate and expand in the false vacuum background. This may lead to the formation of PBHs, if the bubble collisions compress a large amount of energy into a small volume~\cite{Hawking:1982ga,Lewicki:2019gmv,Jung:2021mku}, or if some false vacuum remnants form significant overdensities via delayed vacuum decay~\cite{Kodama:1982sf,Hall:1989hr,Kusenko:2020pcg,Liu:2021svg,Hashino:2021qoq,Hashino:2022tcs,He:2022amv,Kawana:2022olo,Lewicki:2023ioy,Gouttenoire:2023naa,Gouttenoire:2023pxh,Salvio:2023blb,Conaci:2024tlc,Banerjee:2023brn,Banerjee:2023vst,Banerjee:2023qya,Baldes:2023rqv,Lewicki:2024ghw,Balaji:2024rvo,Jinno:2023vnr,Flores:2024lng} or trapping fermions~\cite{Lee:1986tr,Gross:2021qgx,Baker:2021nyl,Baker:2021sno,Kawana:2021tde,Huang:2022him,Kawana:2022lba,Lu:2022jnp,Marfatia:2021hcp,Marfatia:2022jiz,Tseng:2022jta,Xie:2023cwi,Acuna:2023bkm,Lewicki:2023mik,Chen:2023oew,Kim:2023ixo,Gehrman:2023qjn,DelGrosso:2024wmy,Borah:2024lml}. While FOPTs might be generic during the thermal history of the Universe~\cite{Mazumdar:2018dfl}, the formation of PBHs during such transitions is highly nontrivial. It typically imposes additional constraints on the temperature, strength, duration, and bubble expansion velocity of the FOPT, as well as the particle content of the underlying model. Depending on scenarios, the PBHs from FOPTs have various mass, spin, and abundance distributions.

In this work, we focus on the mechanism based on the delayed vacuum decay~\cite{Liu:2021svg} (also see a similar scenario in Ref.~\cite{Kodama:1982sf}). During the FOPT, the energy stored in the false vacuum regions is released and converted to radiation energy. Because of the randomness of the vacuum tunneling process, the timing of bubble nucleation varies across Hubble patches. Some patches experience delayed nucleation compared to the average time, and become overdense over time as they have larger vacuum energy fractions, whose density maintains a constant, while the overall Universe is dominated by radiation that dilutes its energy density via expansion. Those patches eventually decay and release vacuum energy to radiation energy, and when the overdensity reaches a certain threshold, the delayed-decay patches collapse into PBHs.

This mechanism has been extensively studied in the literature on its detailed calculations, applications, and phenomenology~\cite{Hashino:2021qoq,Hashino:2022tcs,He:2022amv,Kawana:2022olo,Lewicki:2023ioy,Gouttenoire:2023naa,Gouttenoire:2023pxh,Salvio:2023blb,Conaci:2024tlc,Banerjee:2023brn,Banerjee:2023vst,Banerjee:2023qya,Baldes:2023rqv,Lewicki:2024ghw}. Most studies are agnostic about the underlying particle models, employing the FOPT parameters such as $\alpha$ and $\beta/H_*$ as input parameters. In particular, the exponential nucleation rate approximation $\Gamma(t)\sim e^{\beta t}$ is commonly adopted. In this paper, we focus on the implementation of the mechanism in particle physics models beyond the Standard Model (SM), and for the first time analyze its general relationship with the scalar potential structure of the model. By investigating models that incorporate {\it zero-temperature potential barriers}, we find that such models typically exhibit a U-shaped Euclidean action $S_3(T)/T$, which results in slow FOPTs, conducive to delayed vacuum decay and hence PBH formation. The PBH abundance is very sensitive to the height of the zero-temperature barrier. We also confirm that the $\Gamma(t)\sim e^{\beta t}$ approximation is not suitable due to its reliance on the linear time-expansion of the Euclidean action, which fails to capture the characteristics of a U-shaped action in such models~\cite{Ellis:2018mja}.

To illustrate our findings, we consider two concrete models. The first model utilizes a polynomial potential, serving as a prototype or toy model. The second model is the $\Z_2$-preserving real singlet extension of the SM (dubbed as the $\Z_2$-xSM), representing a realistic scenario concerning the first-order electroweak phase transition. In Section~\ref{sec:algorithm}, we describe our calculation framework in detail, and also compare different approximation methods for the nucleation rate. Subsequently, we apply this framework to analyze the polynomial potential in Section~\ref{sec:polynomial} and the $\Z_2$-xSM model in Section~\ref{sec:xSM}, presenting our main results. The conclusion will be given in Section~\ref{sec:conclusion}.

\section{Framework for calculation}\label{sec:algorithm}

\subsection{Deriving the PBH profiles}\label{subsec:calculation}

In the early Universe, the scalar potential's shape changes with temperature. Initially, the Universe occupies the minimum of the potential, known as the vacuum. However, below the critical temperature, the potential develops another deeper minimum, which becomes the true vacuum. The Universe remains stuck in the initial false vacuum due to a barrier that prevents a smooth transition to the global minimum. Thermal tunneling allows the Universe to transition to the true vacuum with a probability per unit volume and per unit time~\cite{Linde:1981zj}
\be\label{GammaT}
\Gamma(T)\sim T^4\left(\frac{S(T)}{2\pi}\right)^{3/2}e^{-S(T)},
\ee
where $S(T)=S_3(T)/T$, and $S_3(T)$ is the $O(3)$-symmetric Euclidean action evaluated from the scalar potential. As the temperature $T$ depends on cosmic time $t$, we can express the action and decay rate as $S(t)=S\left(T(t)\right)$ and $\Gamma(t)=\Gamma\left(T(t)\right)$, respectively. Denoting $t_c$ as the time corresponding to the critical temperature $T_c$ where the two vacua are degenerate, we have $S(t_c)=\infty$ and $\Gamma(t_c)=0$.

We now describe the evolution of the Universe in a FOPT. For a given Hubble patch, according to the first Friedmann equation, the Hubble constant is given by
\be\label{1st_Friedmann}
H(t)=\frac{\dot a(t)}{a(t)}=\sqrt{\frac{8\pi}{3M_{\rm Pl}^2}\rho(t)},
\ee
where $M_{\rm Pl}=1.22\times10^{19}$ GeV is the Planck scale, $a(t)$ is the scale factor, and $\rho(t)=\rho_r(t)+\rho_v(t)$ is the energy density that can be decomposed into the radiation and vacuum components
\be
\rho_r(t)=\frac{\pi^2}{30}g_*T^4(t),\quad \rho_v(t)=F(t)\Delta V(t),
\ee
where the factor $g_*$ is the number of effective degrees of freedom, and $\Delta V(t)$ is the energy density difference between the false and true vacua, which is 0 for $t\leqslant t_c$ and positive for $t>t_c$. The false vacuum volume fraction is defined as
\be\label{Ft}
F(t)=\begin{dcases}~1,&\text{if $t<t_d$;}\\
~\exp\left\{-\frac{4\pi}{3}\int_{t_d}^{t}\d t'\Gamma(t')a^3(t')\left[\int_{t'}^t\d t''\frac{v_w}{a(t'')}\right]^3\right\},&\text{if $t>t_d$.}
\end{dcases}
\ee
where $t_d$ is the time that the Hubble patch starts to nucleate, and $v_w$ is the bubble expansion velocity.

As the FOPT proceeds, $F(t)$ decreases from 1 to 0, and the vacuum energy is converted to radiation. This can be clearly seen from the variation of the second Friedmann equation,
\be\label{2nd_Friedmann}
\frac{\d \rho_r}{\d t}+4H\rho_r=-\frac{\d\rho_v}{\d t}.
\ee
Given $t_d$, with the initial condition $\rho_r(t_c)=\pi^2g_*T_c^4/30$ and $\rho_v(t_c)=0$, one is able to resolve Eqs.~(\ref{1st_Friedmann}), (\ref{Ft}), and (\ref{2nd_Friedmann}) consistently by iteration, and obtain the evolutions of the physical observables, such as $T(t)$, $\rho_{r,v}(t)$, $H(t)$, and $F(t)$, etc. As mentioned before, $t_d$ varies in different patches since vacuum tunneling is probabilistic. For simplicity, we categorize the Hubble patches into two types: normal ones and delayed-decay ones. Normal patches initiate nucleation after the Universe cools to the critical temperature, i.e. $t_d=t_c$, which is the most prevalent case. Delayed-decay patches, on the other hand, have a late nucleation time, i.e. $t_d>t_c$. We use the subscripts ``out'' and ``in'' to label the two types of patches and the corresponding physical observables, respectively. The normal patch can be resolved once the underlying particle model is set, while the delayed-decay patch is determined by one extra parameter $t_d$.

The delayed-decay patches form overdensities with respect to to normal patches because the latter receive more radiation contributions, and $\rho_r\propto a^{-4}$ redshifts rapidly while $\rho_v\propto a^0$ remains a constant. The overdensity is defined as
\be\label{delta}
\delta(t) = \frac{\rho_{\rm in}(t)}{\rho_{\rm out}(t)}-1=\frac{\rho_{r,\rm in}(t)+\rho_{v,\rm in}(t)}{\rho_{r,\rm out}(t)+\rho_{v,\rm out}(t)}-1.
\ee
As time $t$ increases from $t_c$, $\delta(t)$ increases from 0 to a maximum value, $\delta_{\rm max}$, and then decreases back to zero as $t\to\infty$. If $\delta_{\rm max}$ exceeds a threshold value $\delta_c$, the delayed-decay patch collapses into a PBH at $t_{\rm pbh}$ satisfying $\delta(t_{\rm pbh})=\delta_c$. In this paper, we take $\delta_c=0.45$ as the PBH formation criterion~\cite{Musco:2004ak,Harada:2013epa}. This criterion can be obtained under the assumption that the Universe is radiation dominant and that the overdensity region is spherically symmetric.
If the overdensity region occurs due to vacuum energy, an inflationary expanding baby Universe is realized there~\cite{Berezin:1982ur,Blau:1986cw,Garriga:2015fdk,Deng:2020mds}, and the region appears to the outside observer as a PBH. As it has not been numerically shown what value of $\delta_{c}$ should be taken in such situations, we take the well-known value $\delta_{c} =0.45$ as a benchmark in our study.

We can establish a mapping between $t_d$ and $\delta_{\rm max}$. Generally, $\delta_{\rm max}$ increases with $t_d$ since a patch with later vacuum decay forms a larger energy contrast. However, the probability of ``a patch remains in the false vacuum until $t_d$'' decreases rapidly as $t_d$ increases. Thus, we infer that the smallest value of $t_d$ allowing for collapse dominates the abundance of PBHs. This corresponds to the value of $t_d$ where $\delta_{\rm max}=\delta_c$, and the moment when this maximum value is reached defines the PBH formation time $t_{\rm pbh}$, specifically $\delta(t_{\rm pbh})=\delta_{\rm max}=\delta_c$. By this procedure, we can determine $t_d$ and $t_{\rm pbh}$ simultaneously (note $t_{\rm pbh}>t_d$). The resultant PBH mass is
\be
m_{\rm pbh}\approx\gamma V_{H,\rm in}(t_{\rm pbh})\rho_{\rm in}(t_{\rm pbh})\approx\gamma\times\frac{4\pi}{3} H_{\rm in}^{-3}(t_{\rm pbh})\times\frac{3M_{\rm Pl}^2}{8\pi}H_{\rm in}^2(t_{\rm pbh}),
\ee
where $V_H(t_{\rm pbh})=(4\pi/3)H_{\rm in}^{-3}(t_{\rm pbh})$ is the Hubble volume. The factor $\gamma$ represents the ratio between the PBH mass and the Hubble mass. We here take $\gamma\approx0.2$~\cite{Carr:1975qj}.

To determine the abundance of PBHs, it is necessary to evaluate the probability of a patch nucleating at time $t_d$. That is to find a region with a Hubble volume $(4\pi/3)H_{\rm in}^{-3}(t_{\rm pbh})$ that remains bubble-free until $t_d$. Within the time interval $t_c<t<t_{\rm pbh}$, this region has a volume of $(4\pi/3)[a_{\rm in}(t)/a_{\rm in}(t_{\rm pbh})]^3H_{\rm in}^{-3}(t_{\rm pbh})$, and the probability of no bubble formation in this region during the time interval $[t,t+\d t]$ is given by
\be
\d P=1-\frac{4\pi}{3}\frac{a_{\rm in}^3(t)}{a_{\rm in}^3(t_{\rm pbh})}\frac{\Gamma_{\rm in}(t)}{H_{\rm in}^{3}(t_{\rm pbh})}\d t\approx\exp\left\{-\frac{4\pi}{3}\frac{a_{\rm in}^3(t)}{a_{\rm in}^3(t_{\rm pbh})}\frac{\Gamma_{\rm in}(t)}{H_{\rm in}^{3}(t_{\rm pbh})}\d t\right\}.
\ee
Consequently, the probability of no bubble nucleation during the time interval $t\in[t_c,t_d]$ is given by
\be
P(t_d)=\prod\d P=\exp\left\{-\frac{4\pi}{3}\int_{t_c}^{t_d}\d t\frac{a_{\rm in}^3(t)}{a_{\rm in}^3(t_{\rm pbh})}H_{\rm in}^{-3}(t_{\rm pbh})\Gamma_{\rm in}(t)\right\}.
\ee
This expression demonstrates that $P(t_d)$ is exponentially dependent on $t_d$, emphasizing the significance of setting the PBH formation time and mass through $\delta(t_{\rm pbh})=\delta_{\rm max}=\delta_c$. This lower limit on $t_d$ governs the PBH density. 

With $P(t_d)$ in hand, we can calculate $\rho_{\rm pbh}$. Suppose the numbers of the delayed-decay and normal Hubble patches are $N_{\rm in}$ and $N_{\rm out}$, respectively, and then the energy density of PBH is
\be\label{rhopbhbare}
\rho_{\rm pbh}=\frac{N_{\rm in}m_{\rm pbh}}{N_{\rm out}V_{H,\rm out}(t_{\rm pbh})+N_{\rm in}V_{H,\rm in}(t_{\rm pbh})},
\ee
where $V_H$ is again the Hubble volume. The delayed-decay probability $P(t_d)$ implies
\be
P(t_d)=\frac{N_{\rm in}}{N_{\rm in}+N_{\rm out}},
\ee
and substituting this relation back to \Eq{rhopbhbare} yields
\be
\rho_{\rm pbh}=\frac{P(t_d)\cdot m_{\rm pbh}}{\left(1-P(t_d)\right)V_{H,\rm out}+P(t_d)V_{H,\rm in}}\approx P(t_d)\frac{m_{\rm pbh}}{V_{H,\rm out}},
\ee
where the approximate equality holds because $P(t_d)\ll1$. The present-day energy density of PBHs is given by $\rho_{\rm pbh}^0=\rho_{\rm pbh}s_0/s$, where $s\approx2\pi^2g_{*s}T_{\rm out}^3(t_{\rm pbh})/45$ and $s_0\approx2891.2~{\rm cm^{-3}}$~\cite{ParticleDataGroup:2022pth} the entropy at PBH formation and today, respectively. The relic abundance of PBHs can be expressed as $\Omega_{\rm pbh}=\rho_{\rm pbh}^0/\rho_0$, where $\rho_0=(3M_{\rm Pl}^2/8\pi)H_0^2$ with $H_0\approx 67.4~{\rm km/(s\cdot Mpc)}$ the current Hubble constant~\cite{ParticleDataGroup:2022pth}. We further define the PBH fraction of dark matter as
\be
f_{\rm pbh}=\frac{\Omega_{\rm pbh}}{\Omega_{\rm dm}},
\ee
where $\Omega_{\rm dm}h^2\approx0.12$ is the dark matter relic abundance.

We have described our framework to determine the formation time, mass, and relic abundance of PBHs. The formalism applies to general particle models with various shapes of potentials. Our calculation disregards the FOPT reheating effect, assuming a constant product of temperature ($T$) and scale factor ($a$). This simplification may not hold for ultra-supercooled FOPTs with the parameter $\alpha\gg1$, where the Universe's temperature can be increased by a factor of approximately $(1+\alpha)^{1/4}$ after the transition. However, our analysis focuses only on the mild-supercooled case when $\alpha$ is moderate, making the reheating effect negligibly small. More discussions on calculations involving very strong FOPTs can be found in Refs.~\cite{Lewicki:2023ioy,Gouttenoire:2023naa}.

\subsection{The exponential approximation and beyond}\label{subsec:expansions}

Once the underlying particle model is built, one is able to derive $S(t)$ and also $\Gamma(t)$, and proceed the calculation as described in Section~\ref{subsec:calculation}. To simplify the calculation, most previous studies have chosen to expand the action at a specific moment $t_*$,
\be\label{S_expansion}
S(t)\approx S(t_*)-\beta (t-t_*)+\frac{\zeta^2}{2}(t-t_*)^2+\cdots,
\ee
and have written down
\be
\Gamma(t)\approx \Gamma_*e^{\beta (t-t_*)-\frac{\zeta^2}{2}(t-t_*)^2+\cdots},
\ee
where $T_*$ is the temperature at $t_*$, and $\Gamma_*=T_*^4\left(S(t_*)/(2\pi)\right)^{3/2}$. Here the Taylor expansion coefficients are
\be
\beta=-\frac{\d S}{\d t}\Big|_*,\quad \zeta^2=\frac{\d^2 S}{\d t^2}\Big|_*,
\ee
or they can be rewritten in the dimensionless forms and transferred into the derivatives on temperature, which are more explicitly derived from the particle model:
\be\label{beta_zeta}
\frac{\beta}{H_*}=T_*\frac{\d S}{\d T}\Big|_*,\quad \left(\frac{\zeta}{H_*}\right)^2=T_*^2\frac{\d^2S}{\d T^2}\Big|_*+\left(1+\frac{\d\ln H}{\d\ln T}\Big|_*\right)\frac{\beta}{H_*},
\ee
where $H_*$ is the Hubble constant at $t_*$, and $S(T)=S_3(T)/T$.

Regarding the expansion \Eq{S_expansion}, the commonly used approach is to consider only the linear $\beta$-term, leading to the exponential nucleation rate approximation $\Gamma(t)\approx\Gamma_*e^{\beta(t-t_*)}$, where $\beta^{-1}$ is interpreted as the time scale of the phase transition. Therefore $\beta/H_*$ can be understood as the ratio of Hubble time scale to FOPT duration. Besides $\beta/H_*$, there is another important parameter $\alpha$ describing the strength of the FOPT, defined as
\be
\alpha=\left(\Delta V(t_*)-T(t_*)\frac{\partial\Delta V}{\partial T}\Big|_*\right)/\rho_r(t_*),
\ee
i.e. the ratio of phase transition latent heat to the radiation energy density, evaluated in the normal patches. The parameters $\alpha$ and $\beta/H_*$, together with the transition temperature $T_*$ and bubble wall velocity $v_w$, can be used to calculate other observables of the FOPT dynamics, such as the stochastic gravitational wave (GW) spectrum~\cite{Mazumdar:2018dfl,Caprini:2019egz,Guo:2021qcq,Athron:2023xlk}.

As for the expansion moment $t_*$, there are typically two choices. One is the nucleation time $t_n$, the moment that the bubble number in the normal patch reaches 1, i.e.
\be
N(t_n)=\int_{t_c}^{t_n}\d t'\frac{\Gamma(t')}{H^3(t')}=1.
\ee
The other is the percolation time $t_{p}$, which is the moment when the bubbles in the normal patch form an infinite connecting cluster, i.e., $F(t_p)=0.71$~\cite{rintoul1997precise}. In general $t_p>t_n$, and when the FOPT is prompt ($\beta/H_*\gg1$) and moderate ($\alpha\lesssim1$), $t_p\approx t_n$. However, if the FOPT is very strong or lasts for a long time, $t_p$ differs significantly from $t_n$. In that case, it is proposed to adopt $t_*=t_p$, and replace $\beta/H_*$ with $(8\pi)^{1/3}v_w/(H_*\bar R)$ in the GW calculation, where $\bar R$ is the mean separation of bubbles~\cite{Ellis:2018mja,Ellis:2019oqb,Wang:2020jrd}, which can be adopted as $[n_b(t_*)]^{-1/3}$, with
\be
n_b(t)=\int_{t_c}^{t}\d t'F(t')\Gamma(t')\frac{a^3(t')}{a^3(t)},
\ee
the bubble number density in the normal patches. Furthermore, one should also check the FOPT completion condition~\cite{Turner:1992tz,Ellis:2018mja}
\be\label{didt}
3H(t_p)+\frac{\d\ln F(t)}{\d t}\Big|_p<0,
\ee
such that the physical volume of the false vacuum is decreasing at percolation.

The exponential approximation is widely used in FOPT research and is generally effective. It is also commonly employed in the study of PBH formation from delayed vacuum decay~\cite{Liu:2021svg,Hashino:2021qoq,Hashino:2022tcs,He:2022amv,Kawana:2022olo,Gouttenoire:2023naa,Gouttenoire:2023pxh,Salvio:2023blb,Conaci:2024tlc,Banerjee:2023brn,Banerjee:2023vst,Banerjee:2023qya,Baldes:2023rqv,Lewicki:2024ghw,Balaji:2024rvo}. However, PBH formation predominantly occurs in the small $\beta/H_*$ region, which corresponds to slow FOPTs with prolonged durations. For instance, previous studies indicate $\beta/H_*\lesssim10$ for abundant PBH formation~\cite{Liu:2021svg,Lewicki:2023ioy,Gouttenoire:2023naa}. When the first-order expansion coefficient $\beta$ of $S(t)$ is small, it becomes necessary to check if truncating the expansion \Eq{S_expansion} at linear order provides a satisfactory approximation. This can be determined by examining whether the ratio of the second-order expansion to the first-order one, $\left|[\zeta^2(t-t_*)^2/2]/[\beta (t-t_*)]\right|$, is much less than 1 during the FOPT. This is equivalent to checking
\be\label{linear_condition}
\frac12\left|\left(1+\frac{\d\ln H}{\d\ln T}\Big|_*+\frac{\d\ln S'}{\d \ln T}\Big|_*\right)H_*(t-t_*)\right|\ll1,
\ee
where $S'$ represents $\d S/\d T$.

The factor $H_*(t-t_*)$ roughly measures the duration of FOPT over the Hubble time scale, and according to the calculations in Appendix~\ref{app:linear_expansion}, it can be replaced by $\mO(10)\times(\beta/H_*)^{-1}$. In addition, we can generally assume $H\propto T^n$ and $S\propto T^r$ near $T_*$, and then \Eq{linear_condition} reduces to
\be\label{reduced_linear_con}
|n+r|\ll \mO(0.2)\times\frac{\beta}{H_*}.
\ee
PBH formation requires the right-hand side (r.h.s.) of \Eq{reduced_linear_con} to be $\sim0.2\times10\sim2$. However, in ultra-supercooled and mild FOPTs with vacuum and radiation domination, the left-hand side (l.h.s.) typically takes values of $n\approx0$ and $n\approx2$ respectively, while a realistic particle model usually has $r>1$. Consequently, it is challenging to satisfy the exponential approximation condition in the parameter space favored by the PBH mechanism. If the l.h.s. is not sufficiently small, including additional terms in the expansion, such as the quadratic $\zeta$-term~\cite{Lewicki:2023ioy}, becomes necessary.

In our study, we directly compare the results obtained using three different methods for $\Gamma(t)$: (1) the full expression, (2) the first-order expansion with only the $\beta$-term (exponential approximation), and (3) the second-order expansion with both the $\beta$- and $\zeta$-terms. We find that the first-order expansion is inadequate, and it results in significant differences in PBH mass and abundance compared to the results obtained from the full expression case. By including the quadratic term, the results are much improved and closer to those of the full expression case, although visible differences still persist, particularly for relatively mild FOPTs. Our results demonstrate the importance of utilizing the full expressions for $\Gamma(t)$ in the PBH formation calculation in realistic models.

\section{The model with a polynomial potential}\label{sec:polynomial}

\subsection{Model description and PBH formation}

Let $\phi$ be the scalar that triggers the FOPT, and $V(\phi,T)$ its potential in the thermal bath of the early Universe at temperature $T$, we parametrize the potential as
\be\label{TBR}
V(\phi,T)=\frac12\left(\frac{\mu_3 w-m_\phi^2}{2}+c\,T^2\right)\phi^2-\frac{\mu_3}{3}\phi^3+\frac{m_\phi^2}{8w^2}\left(1+\frac{\mu_3w}{m_\phi^2}\right)\phi^4,
\ee
and restrict the value of $\mu_3$ to be within the range
\be \label{mu3_region}
\frac{m_\phi^2}{w}<\mu_3<\frac{3m_\phi^2}{w}.
\ee
At $T=0$, $V(\phi,0)$ has two minima at $\phi=0$ and $w$, and the latter is the true vacuum. The $\phi$ boson mass square is $m_\phi^2=\d^2V(\phi,0)/\d\phi^2\big|_{\phi=w}$. Between the two minima, there is a local maximum at $\phi=w_b$, corresponding to the zero-temperature barrier
\be
V(w_b,0)=\frac{m_\phi^2w^2}{24}\left(3+\frac{\mu_3w}{m_\phi^2}\right)\left(1-\frac{2}{1+w\mu_3/m_\phi^2}\right)^3.
\ee
It exists when $\mu_3>m_\phi^2/w$. If we increase $\mu_3$, the barrier height increases accordingly, and $V(w,0)$ is also shifted. When $\mu_3$ reaches $3m_\phi^2/w$, the $\phi=w$ vacuum becomes degenerate with the $\phi=0$ one. Therefore, we consider $\mu_3$ in the range in \Eq{mu3_region} to ensure there is a barrier at zero-temperature, and $\phi=w$ is the true vacuum.

\begin{figure}
\begin{center}
\includegraphics[scale=0.35]{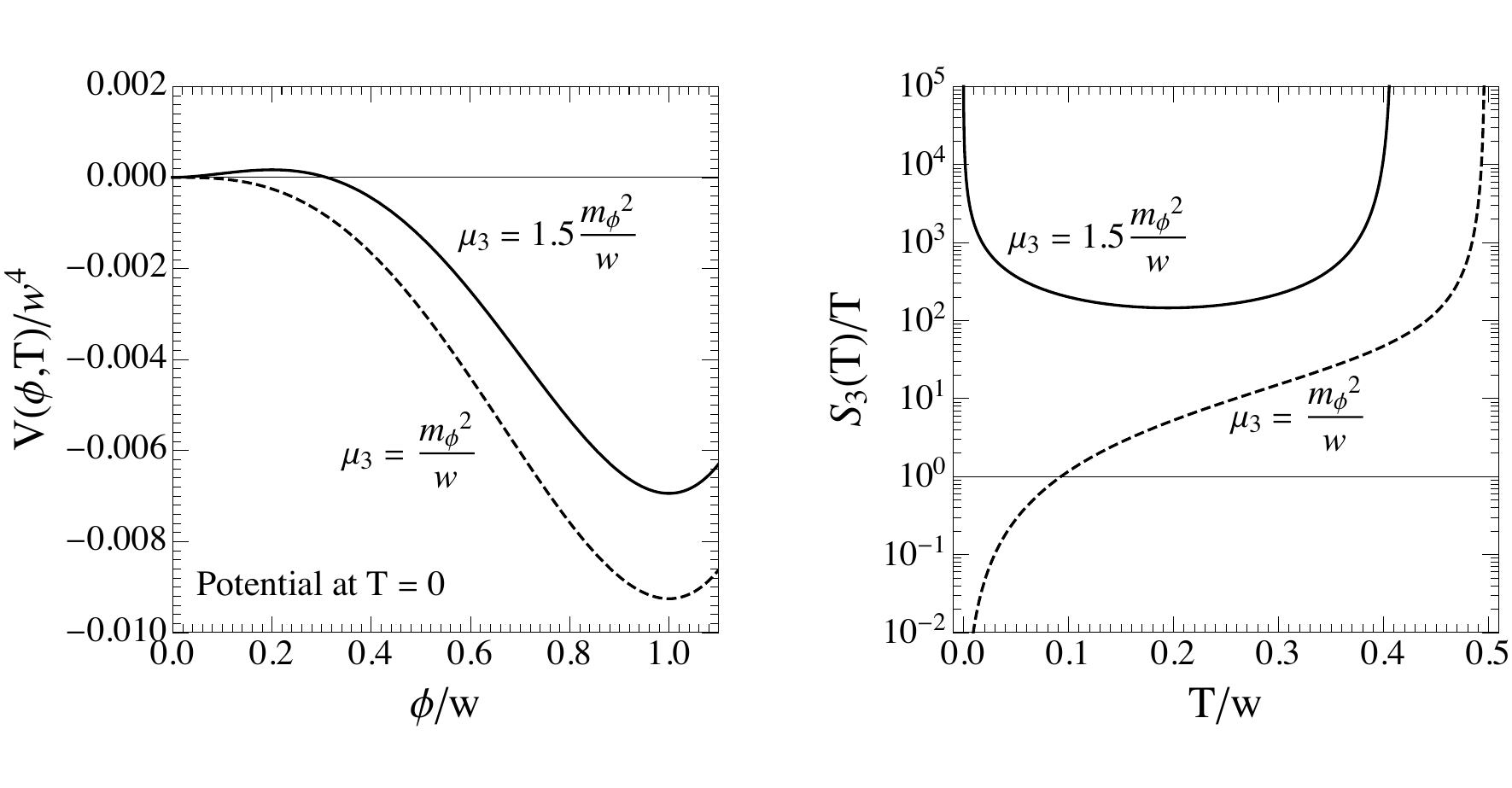}
\caption{Left: the zero-temperature potentials. Right: the $S_3(T)/T$ curves. $m_\phi=300$ MeV, $w=900$ MeV, and $c=0.11$, while $\mu_3$ takes $m_\phi^2/w=100$ MeV or $1.5m_\phi^2/w=150$ MeV.}
\label{fig:potential}
\end{center}
\end{figure}

The finite temperature correction to the potential is described by the $c$-term in \Eq{TBR}, where we only include the leading $T^2$-terms of the thermal integrals coming from light degrees of freedom. The subdominant $-c'T\phi^3$ term from the bosonic degrees of freedom is omitted. The Universe stays in the false vacuum $\phi=0$ initially and it acquires a probability of decaying to the true vacuum when $T<T_c$. The decay rate $\Gamma(T)\sim e^{-S_3(T)/T}$ is defined in \Eq{GammaT}. At $T=T_c$, the two vacua are degenerate and hence $S_3(T_c)=\infty$; when the Universe cools down from $T_c$, $S_3(T)/T$ also decreases. If there exists a zero-temperature barrier, then $S_3(T)$ approaches a finite value for $T\to0$, resulting in $\lim_{T\to0}S_3(T)/T=\infty$. Therefore, in such a case we have a U-shaped $S(T)=S_3(T)/T$ curve, which efficiently suppresses the vacuum decay rate and hence results in a long FOPT. To illustrate this, we adopt $m_\phi=300~{\rm MeV}$, $w=900~{\rm MeV}$, and $c=0.11$ as a benchmark, and test two different $\mu_3$ values: $m_\phi^2/w=100$ MeV and $1.5m_\phi^2/w=150$ MeV. The former corresponds to vanishing barrier at zero-temperature and hence $S_3(T)/T\to0$ when $T\to0$, while the latter corresponds to a barrier that results in a U-shaped $S_3(T)/T$ curve, as clearly illustrated in Fig.~\ref{fig:potential}. For the $S_3(T)/T$ calculation, we have used the semi-analytical expression~\cite{Dine:1992wr}
\be\label{S3T_numerical}
\frac{S_3(T)}{T}\approx\frac{123.48(\mu^2+c\,T^2)^{3/2}}{2^{3/2}T\mu_3^2}f\left(\frac{9(\mu^2+c\,T^2)\lambda}{2\mu_3^2}\right),
\ee
where
\be
\mu^2=\frac{\mu_3 w-m_\phi^2}{2},\quad \lambda=\frac{\mu_3w+m_\phi^2}{2w^2},\quad f(u)=1+\frac{u}{4}\left[1+\frac{2.4}{1-u}+\frac{0.26}{(1-u)^2}\right].
\ee
The approximation works well when $u\in[0,1]$. More discussions on the FOPT dynamics on the U-shaped $S_3(T)/T$ can be found in Refs.~\cite{Megevand:2016lpr,Cai:2017tmh}, while here we only focus on the impacts on PBH formation.

\begin{figure}
\begin{center}
\includegraphics[width=0.99\textwidth]{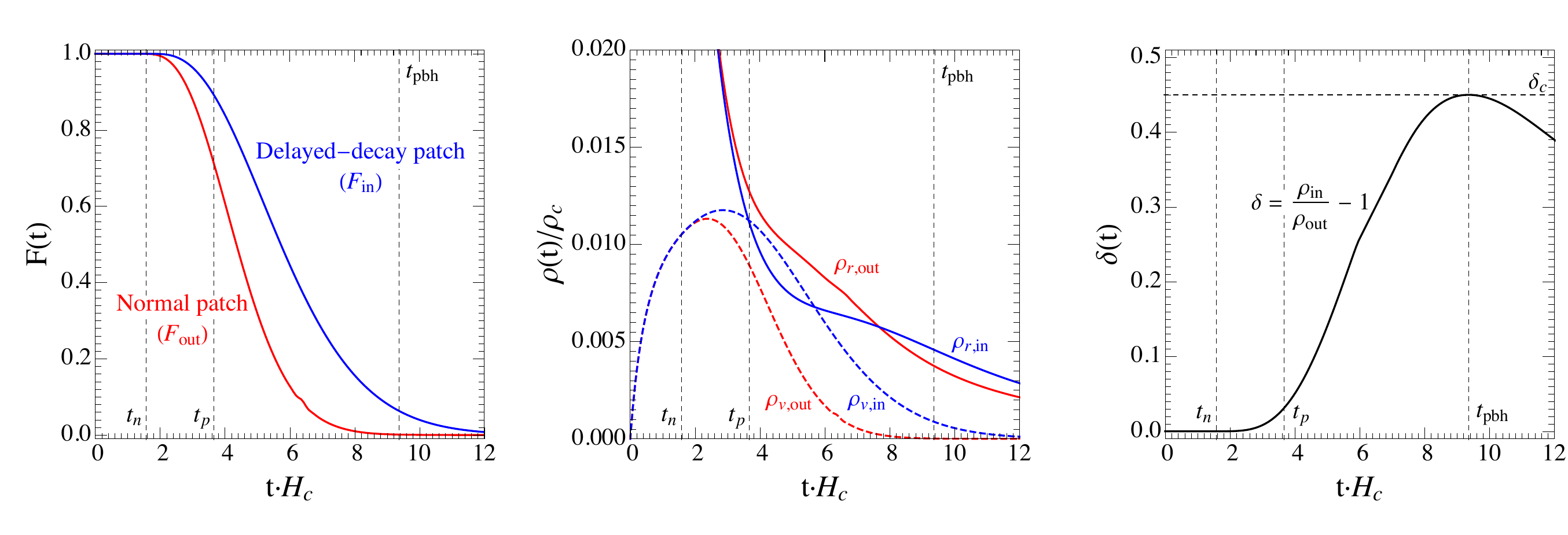}
\caption{The FOPT and PBH profiles for $m_\phi=300$ MeV, $w=900$ MeV, $c=0.11$, and $\mu_3=154.1$ MeV. $t_n$, $t_p$, and $t_{\rm pbh}$ are the nucleation, percolation, and PBH formation time, respectively. $H_c$ and $\rho_c$ denote the Hubble constant and radiation energy density at $t_c$, respectively. Left: the false vacuum fraction $F(t)$. Middle: the energy components $\rho_{r,v}(t)$. Right: the overdensity $\delta(t)$.}
\label{fig:profiles}
\end{center}
\end{figure}

For a given set of $\{m_\phi, w, \mu_3, c\}$, we apply the methodology outlined in Section~\ref{subsec:calculation} to obtain the FOPT profiles and calculate the PBH formation. We set $v_w = 1.0$ as a fixed parameter. An illustrative example is presented in Fig.~\ref{fig:profiles} with $\mu_3 = 154.1$ MeV. In the left panel, we depict the evolution of the false vacuum fraction $F(t)$. From the figure it can be clearly seen that delayed-decay patches initiate nucleation at later times compared to normal patches. Consequently, this results in distinct energy component evolutions, as shown in the middle panel. In both cases, the vacuum energy $\rho_v(t)$ is converted into radiation energy $\rho_r(t)$; however, delayed-decay patches exhibit higher values of $\rho_r(t)$ at later times. The right panel displays the overdensity $\delta(t)$, which progressively increases with time until it reaches the maximum value $\delta_{\rm max}=\delta_c$. This defines the PBH formation time $t_{\rm pbh}=4.70\times10^{-5}$ s. The corresponding PBH mass is $m_{\rm pbh}=2.72\times10^{33}\,{\rm g} \, =1.37\,M_{\odot}$, where $M_{\odot}$ is the solar mass. For reference, the nucleation time $t_n$ and percolation time $t_p$ of the normal patches are also indicated in the figures, and we can see that $t_{\rm pbh}>t_p>t_n$, which is a typical feature of this PBH mechanism.

\begin{figure}
\begin{center}
\includegraphics[scale=0.4]{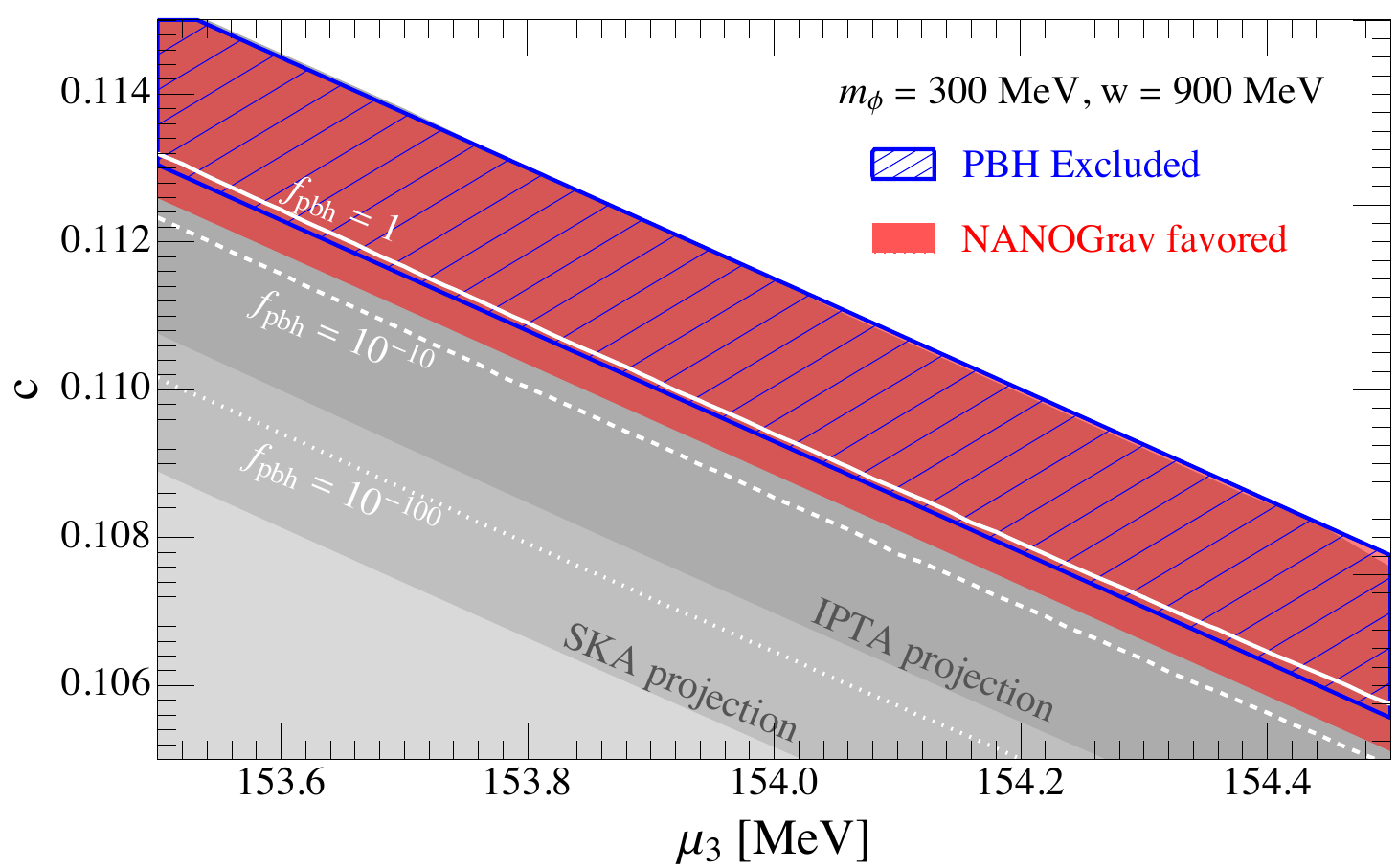}
\caption{The scan result in the $c$-$\mu_3$ plane for $m_\phi=300$ MeV and $w=900$ MeV. The white region cannot realize the FOPT. The white solid, dashed, and dotted lines represent $f_{\rm pbh}=1$, $10^{-10}$, and $10^{-100}$, respectively. The red, gray, and lighter gray regions correspond to the parameter space that produces GWs favored by the NANOGrav excess, reached by future IPTA and SKA, respectively. The blue mesh region is excluded by microlensing experiments.}
\label{fig:2d}
\end{center}
\end{figure}

In Fig.~\ref{fig:2d}, we present the scan results in the $c$-$\mu_3$ plane, with fixed values of $m_\phi=300$ MeV and $w=900$ MeV. By varying $\mu_3$ and $c$ around the center point at $\mu_3=154$ MeV and $c=0.11$, we explore the parameter space for FOPT and PBH formation. 
We observe that the PBH fraction $f_{\rm pbh}$ is highly sensitive to these parameters, exhibiting a narrow band where it varies from $10^{-100}$ (the white dotted line) to $\sim10^7$ (the upper edge of the color shaded region). This variation occurs within a small range, with $\mu_3$ changing by only about $\sim0.6\%$ and $c$ changing by approximately $\sim9\%$. Therefore, $f_{\rm pbh}$ is most sensitive to $\mu_3$, which describes the height of the zero-temperature potential barrier. This can be understood as the duration of FOPT is determined by the barrier, and $f_{\rm pbh}$ is extremely sensitive to this duration, which is a main feature of the PBH formation mechanism through delayed vacuum decay. The resulting PBHs form at around $10^{-5}$ s after the Big Bang, and have a mass of $\sim10^{33}\, {\rm g}\, (\sim M_{\odot})$. Within this mass region, $f_{\rm pbh}$ is constrained to be $\lesssim10^{-2}$ due to microlensing observations~\cite{Carr:2020gox}, and the blue mesh region is already excluded.

\begin{table}\footnotesize\renewcommand\arraystretch{1.5}\centering
\begin{tabular}{|c|c|c|c|c|}\hline
$f_{\rm pbh}$ & $T_*$  & $\alpha$ & $\beta/H_*$ & $\frac{(8\pi)^{1/3}v_w}{H_*\bar R}$ \\ \hline
$10^{-100}$ & $184~{\rm MeV}_n$, $155~{\rm MeV}_p$ & $0.37_n$, $0.99_p$ & $46_n$, $-29_p$ & $3.0_n$, $14_p$ \\ \hline
$10^{-10}$ & $177~{\rm MeV}_n$, $143~{\rm MeV}_p$ & $0.46_n$, $1.5_p$ & $30_n$, $-53_p$ & $3.0_n$, $8.4_p$ \\ \hline
$1$ & $174~{\rm MeV}_n$, $137~{\rm MeV}_p$ & $0.46_n$, $1.9_p$ & $21_n$, $-65_p$ & $3.0_n$, $6.7_p$ \\ \hline
\end{tabular}
\caption{The FOPT parameters corresponding to the scan in Fig.~\ref{fig:2d}. The subscript ``$n$'' and ``$p$'' represent the values derived at nucleation ($t_n$) or percolation ($t_p$), respectively.}
\label{tab:FOPT_parameters}
\end{table}

We evaluate the FOPT parameters, namely $\alpha$, $\beta/H_*$, and $(8\pi)^{1/3}v_w/(H_*\bar R)$, and observe that their contours closely align with those of $f_{\rm pbh}$. To maintain clarity in the figure, we present the corresponding FOPT parameter values for different $f_{\rm pbh}$ in Table~\ref{tab:FOPT_parameters} instead of showing more contours in Fig.~\ref{fig:2d}. We note that $\alpha$ at $t_p$ tends to be larger than at $t_n$ due to the greater dilution of radiation energy, as $t_p>t_n$. We do not have $\alpha\gg1$, and hence the FOPT is not very strong. Regarding parameters related to the FOPT duration, we find that $\beta/H_*$ at $t_n$ and $(8\pi)^{1/3}v_w/(H\bar R)$ at $t_p$ decrease with longer FOPT durations. Conversely, $(8\pi)^{1/3}v_w/(H_*\bar R)$ at $t_n$ remains insensitive to parameter variations, while $\beta/H_*$ at $t_p$ can even become negative. The reasons for these observations will be discussed in the following subsection, and we will demonstrate that $(8\pi)^{1/3}v_w/(H_*\bar R)$ at $t_p$ can measure the ratio of Hubble time to FOPT duration.

The GWs generated by the FOPT are computed using numerical formulae~\cite{Caprini:2019egz}, and their spectral peak falls within the $10-100$ nHz range. This may provide an explanation for the recent excess observed in pulsar timing array (PTA) experiments, such as NANOGrav~\cite{NANOGrav:2023gor}, CPTA~\cite{Xu:2023wog}, EPTA~\cite{EPTA:2023fyk}, and PPTA~\cite{Reardon:2023gzh}. Following the analysis conducted by the NANOGrav collaboration~\cite{NANOGrav:2023hvm}, we define ``favored by the data'' as a GW spectrum that matches the first 14 frequency bins of the NANOGrav-15yr dataset, highlighting this parameter space in red of Fig.~\ref{fig:2d}. The parameter spaces accessible to future PTA experiments, such as IPTA~\cite{Hobbs:2009yy} and SKA~\cite{Weltman:2018zrl}, are depicted in gray and lighter gray shades, respectively. Note that while much of the parameter space capable of explaining the NANOGrav excess has been ruled out by PBH detections, a narrow region remains unexplored, awaiting further investigation. We refrain from delving into the interplay between the PTA data, the FOPT, and PBHs, as the primary focus of this study is the relationship between the structure of particle models and PBH formation.\footnote{For more discussions on the MeV-scale FOPT explanation of the PTA data, see Refs.~\cite{Salvio:2023blb, Han:2023olf, Li:2023bxy, Megias:2023kiy, Fujikura:2023lkn, Zu:2023olm, Athron:2023mer, Jiang:2023qbm, Addazi:2023jvg, Bringmann:2023opz, Madge:2023dxc, Chen:2023bms, He:2023ado, An:2023jxf, Bian:2023dnv, Ellis:2023oxs, Ghosh:2023aum, Xiao:2023dbb} and the references therein. Due to the extensive literature on this topic, we have only included those directly relevant to the June 2023 PTA excess announcement.}

\subsection{Comparison among different expansions of $S(t)$}

In this subsection, we discuss different treatments of the Euclidean action $S(t)$: linear and quadratic approximations, and the full expression. To begin, we need to determine an appropriate expansion point $t_*$. Although the nucleation moment $t_n$ and percolation moment $t_p$ are potential choices, the previous subsection revealed that the region of parameter space where PBH formation is abundant has a negative value of $\beta/H_*$ at $t_p$. This makes it unsuitable for expansion since it corresponds to a decreasing $\Gamma(t)\sim e^{\beta t}$ at the linear level approximation. Therefore, we select $t_n$ as our expansion point.

\begin{figure}
\begin{center}
\includegraphics[scale=0.35]{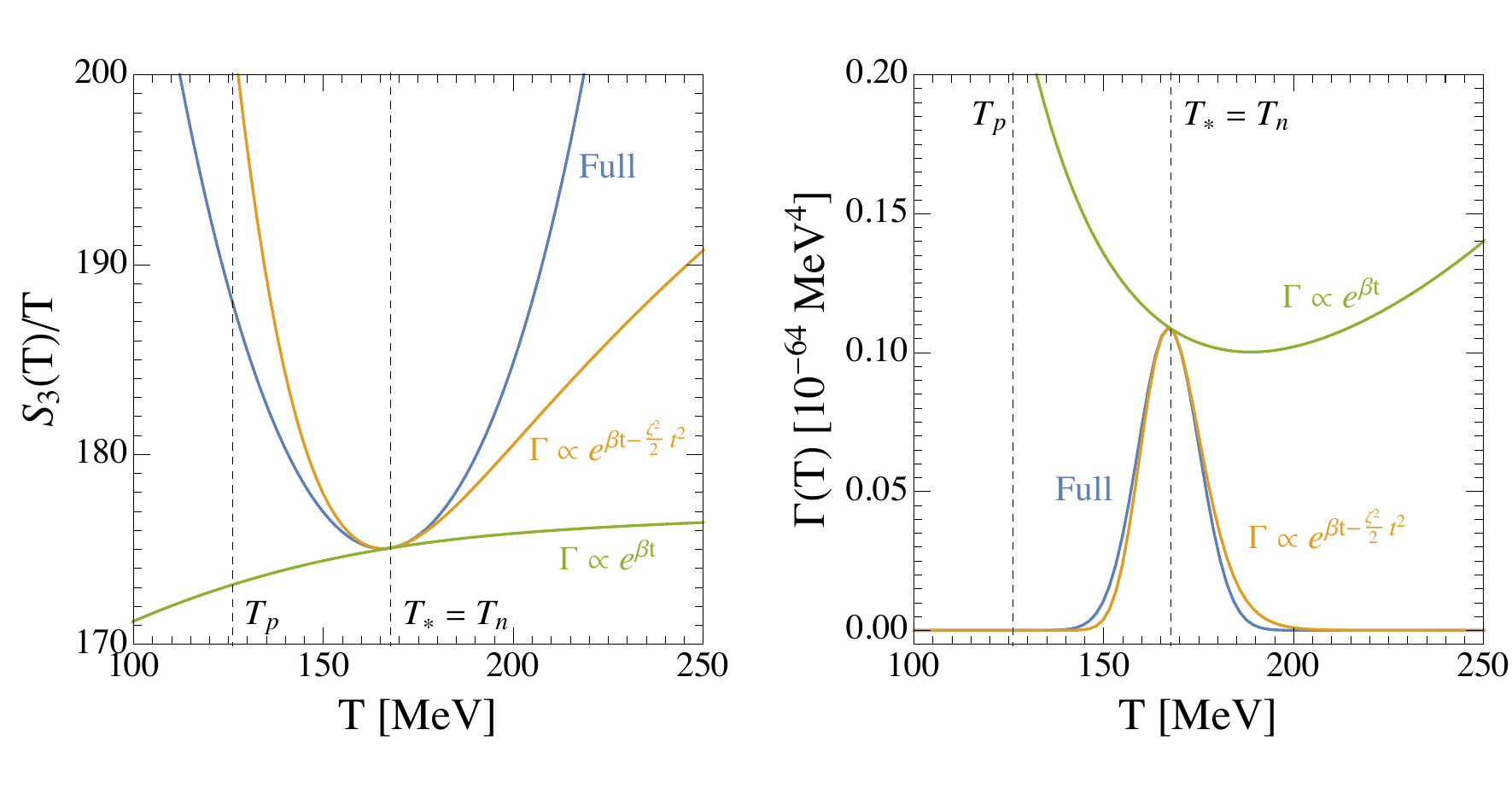}
\caption{The evolutions of $S(T)=S_3(T)/T$ (left) and $\Gamma(T)$ (right), for different treatments: full expression (blue), quadratic expansion (orange), and linear expansion (green). $m_\phi=300$ MeV, $w=900$ MeV, $c=0.11$, and $\mu_3=154.1$ MeV.}
\label{fig:expansions}
\end{center}
\end{figure}

Assuming parameters $m_\phi=300$ MeV, $w=900$ MeV, $c=0.11$, and $\mu_3=154.1$ MeV, Fig.~\ref{fig:expansions} displays the evolutions of the Euclidean action and vacuum decay rate, with the argument switched to temperature for convenience. The left panel shows that the three treatments intersect at $T_n$, but when $T$ deviates from $T_n$, they exhibit distinct shapes, particularly the linear approximation. We now also easily understand why $\beta/H_*$ at $T_p$ is negative: the full $S(T)=S_3(T)/T$ is U-shaped, and hence when the bubbles percolate at late time, $\d S/\d T$ will become negative. The discrepancy in $S(T)$ results in differences in $\Gamma(T)$, as illustrated in the right panel. The exponential nucleation rate approximation fails completely in capturing the peak shape of the decay rate originating from the U-shaped action, whereas the quadratic approximation is much closer to the full expression, albeit with some visible differences. As a result, both the quadratic approximation and the full expression yield $m_{\rm pbh}\sim2.7\times 10^{33} \, {\rm g} \, (\sim 1.36\, M_{\odot})$ and $f_{\rm pbh}\sim 2\times10^6$, while the linear approximation has $m_{\rm pbh}\sim1.6\times10^{34}\,{\rm g}\, (\sim 8.1 \, M_{\odot})$ and $f_{\rm pbh}\sim10^{-6}$.

\begin{figure}
\begin{center}
\includegraphics[scale=0.35]{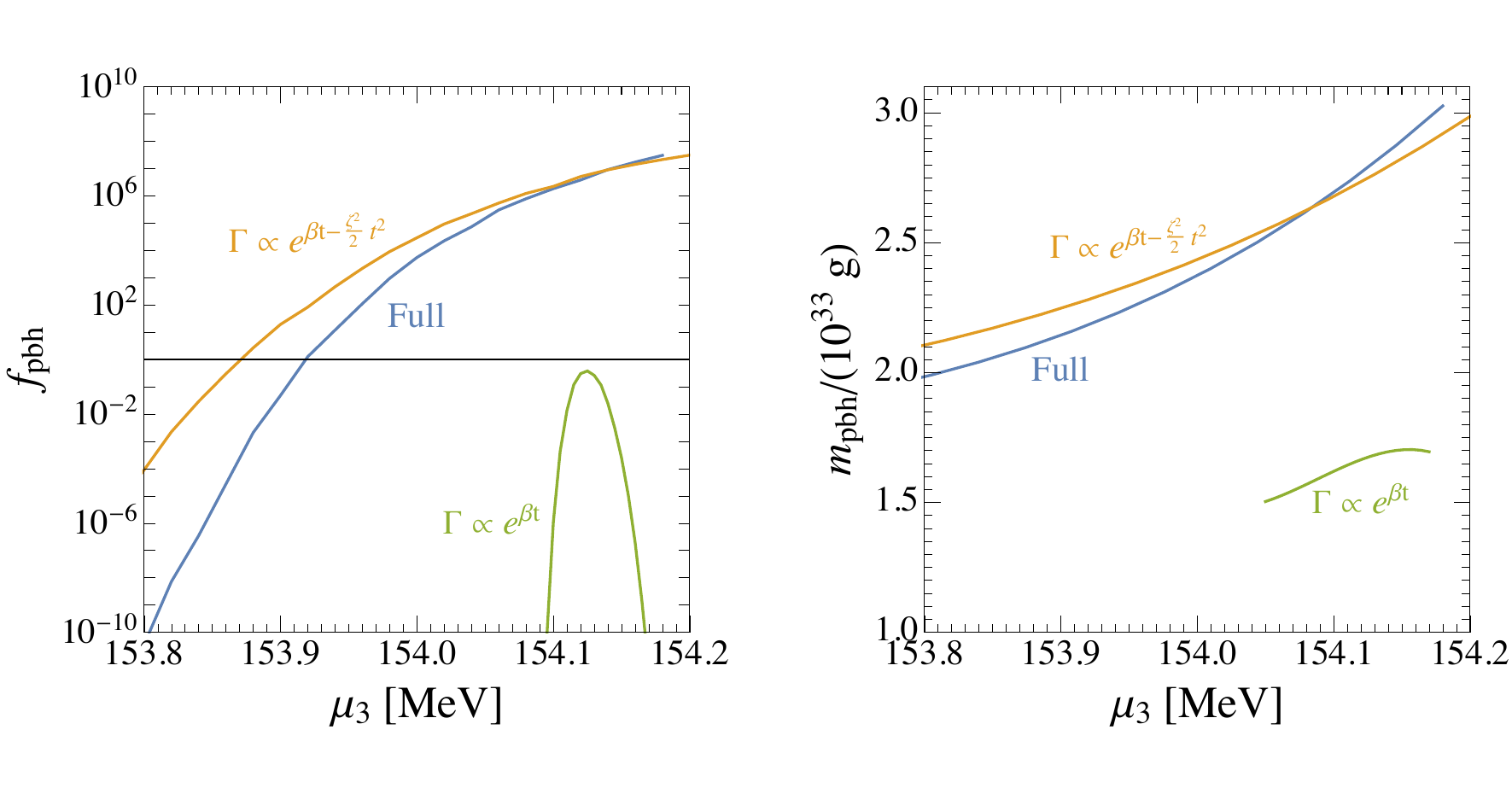}
\caption{The curves of $f_{\rm pbh}$ (left) and $m_{\rm pbh}$ (right), scanning over $\mu_3$ for different treatments of $S(t)$. $m_\phi=300$ MeV, $w=900$ MeV, and $c=0.11$ are fixed.}
\label{fig:mu3_scan}
\end{center}
\end{figure}

The behavior of the linear and quadratic approximations can be understood by assuming $H\propto T^n$ near $T_*$, then
\be
t-t_*=\int_{T}^{T_*}\frac{\d T'}{T'H(T')}\xrightarrow[]{H\propto T^n}\begin{dcases}~\frac{1}{H_*}\frac1n\left[\left(\frac{T_*}{T}\right)^n-1\right],& n\neq0;\\
~\frac{1}{H_*}\log\left(\frac{T_*}{T}\right),& n=0.
\end{dcases}
\ee
and hence \Eq{S_expansion} can also treated as an expansion on temperature $T$, and we can have
\be\label{s_simple}
S(T)-S(T_*)\approx\begin{dcases}~-\frac1n\frac{\beta}{H_*}\left[\left(\frac{T_*}{T}\right)^n-1\right]
+\frac{1}{2n^2}\left(\frac{\zeta}{H_*}\right)^2\left[\left(\frac{T_*}{T}\right)^n-1\right]^2,& n\neq0;\\
~-\frac{\beta}{H_*}\log\left(\frac{T_*}{T}\right)+\frac12\left(\frac{\zeta}{H_*}\right)^2\left[\log\left(\frac{T_*}{T}\right)\right]^2,& n=0,
\end{dcases}
\ee
up to quadratic approximation level. \Eq{s_simple} gives
\be
\frac{\d^2 S}{\d T^2}\Big|_*\approx\frac{1}{T_*^2}\left[-(1+n)\left(\frac{\beta}{H_*}\right)+\left(\frac{\zeta}{H_*}\right)^2\right].
\ee
Since typically $n\in[0,2]$, the linear approximation always has $\d^2S/\d T^2\big|_*<0$, which is qualitatively different from the full U-shaped expression that has $\d^2S/\d T^2\big|_*>0$ as illustrated in Fig.~\ref{fig:expansions}. Such a discrepancy is not pronounced when the FOPT is prompt, as it completes near $T_*$. However, for PBH formation during a prolonged FOPT, the completion temperature deviates from $T_*$. This deviation leads to a significant distinction between the full $S(T)$ and its linear time expansion. Therefore, the exponential approximation $\Gamma(t)\sim e^{\beta t}$ cannot be applied to calculate PBH formation in scalar potentials with zero-temperature barriers when considering delayed vacuum decay. It's worth noting that this argument also applies to more general models with $S(T)\propto T^r$ and $r>1$.

To further explore the differences among the three treatments, we maintain $m_\phi=300$ MeV, $w=900$ MeV, and $c=0.11$, while varying $\mu_3$, and derive the curves of $f_{\rm pbh}$ and $m_{\rm pbh}$ in Fig.~\ref{fig:mu3_scan}. The quadratic approximation exhibits a similar trend to the full expression and provides a close quantitative match in the large $\mu_3$ region. However, for small $\mu_3$, the values of $f_{\rm pbh}$ can differ by several orders of magnitude. In contrast, the linear approximation yields entirely different curves. This discrepancy arises because the decay rate from the linear approximation is generally higher than the other two treatments, as illustrated in the right panel of Fig.~\ref{fig:expansions}. Therefore, it results in faster FOPTs, earlier PBH formation, lower values of $f_{\rm pbh}$, $t_{\rm pbh}$, and $m_{\rm pbh}$.

\begin{figure}
\begin{center}
\includegraphics[width=0.99\textwidth]{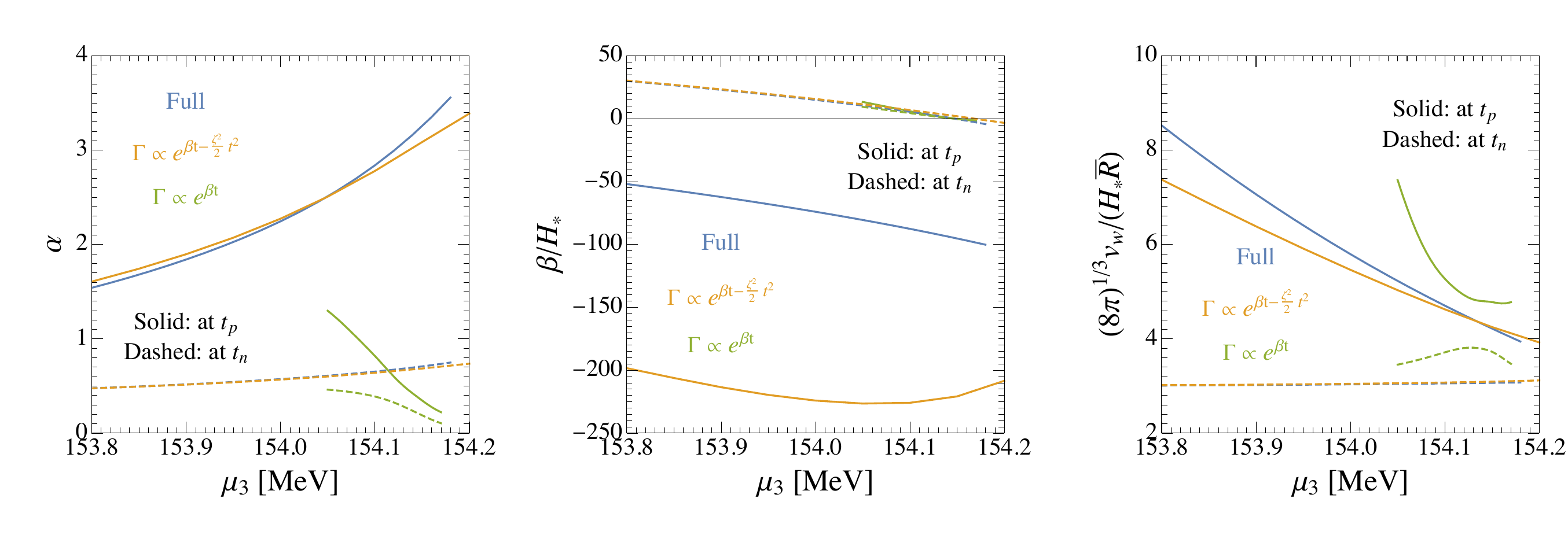}
\caption{The curves of $\alpha_{\rm pbh}$ (left), $\beta/H_*$ (middle), and $(8\pi)^{1/3}v_w/(H_*\bar R)$ (right), scanning over $\mu_3$ for different treatments of $S(t)$. The same parameter setup as Fig.~\ref{fig:mu3_scan}.}
\label{fig:mu3_scan_FOPT}
\end{center}
\end{figure}

We obtain the FOPT parameters $\alpha$, $\beta/H_*$, and $(8\pi)^{1/3}v_w/(H_*\bar R)$ at $t_n$ and $t_p$, and present their dependence on $\mu_3$ in Fig.~\ref{fig:mu3_scan_FOPT}. The left panel demonstrates the FOPT is mild in three different treatments. In the middle panel, we observe that $\beta/H_*$ derived at $t_n$ or $t_p$ becomes negative for sufficiently large $\mu_3$, indicating it cannot be interpreted as a ratio of time scales. Conversely, the right panel demonstrates that $(8\pi)^{1/3}v_w/(H_*\bar R)$ at $t_p$ is positive and decreases with increasing FOPT duration in both the full expression and quadratic approximation cases. Therefore, it serves as a measure of the ratio between the Hubble time and the FOPT time scales.

Before closing this subsection, we emphasize the importance of verifying \Eq{didt} when studying the formation of PBHs resulting from delayed vacuum decay. The FOPTs being considered here have a long duration, necessitating the confirmation of their completion. For instance, in Figs.~\ref{fig:mu3_scan} and~\ref{fig:mu3_scan_FOPT}, the curves representing the full expression evaluation end at $\mu_3\approx154.18$ MeV, indicating that the FOPT cannot be fully accomplished beyond this point. If \Eq{didt} is not satisfied, the physical volume of the false vacuum does not decrease at percolation, even if the volume fraction decreases; consequently, the FOPT remains incomplete~\cite{Turner:1992tz}.

\section{The $\Z_2$-symmetric singlet extension of the SM}\label{sec:xSM}

This section discusses a more complicated and realistic model with double-field FOPT dynamics: the $\Z_2$-symmetric xSM~\cite{McDonald:1993ey,Profumo:2007wc,Espinosa:2011ax}. The scalar sector contains the SM Higgs doublet $H$ and a $\Z_2$-odd real gauge singlet $s$, and the joint potential at finite temperature can be written as
\be\label{VT_sim}
V(h,s,T)\approx\frac{1}{2}\left(\mu_h^2+c_hT^2\right)h^2+\frac12\left(\mu_s^2+c_s T^2\right)s^2+\frac{\lambda_h}{4}h^4+\frac{\lambda_s}{4}s^4+\frac{\lambda_{hs}}{2}h^2s^2,
\ee
where $h$ and $s$ are the neutral Higgs and singlet background field, respectively, and the coefficients
\be
c_h=\frac{3g^2+g'^2}{16}+\frac{y_t^2}{4}+\frac{\lambda_h}{2}+\frac{\lambda_{hs}}{12},\quad c_s=\frac{\lambda_s}{4}+\frac{\lambda_{hs}}{3},
\ee
are derived from the high-temperature expansion of the thermal collections, with $g$, $g'$, and $y_t$ the $SU(2)_L$, $U(1)_Y$ gauge couplings, and the top Yukawa, respectively.\footnote{Here we follow the conventions from Ref.~\cite{Xie:2020wzn}, and the definition of $\lambda_{hs}$ might differ from some references by a factor of 2. It is shown that including the $T^2$-terms yields very similar results compared with the full one-loop calculation~\cite{Xie:2020wzn}.}

We require $\lambda_h>0$, $\lambda_s>0$ and $\sqrt{\lambda_h\lambda_s}+\lambda_{hs}>0$ to ensure the potential bounded from below, and
\be\label{local_minimum}
\mu_h^2<0,\quad \mu_s^2<0,\quad \lambda_h\mu_s^2>\lambda_{hs}\mu_h^2,\quad \lambda_s\mu_h^2>\lambda_{hs}\mu_s^2,\quad -\frac{\mu_h^4}{\lambda_h}<-\frac{\mu_s^4}{\lambda_s},
\ee
such that at $T=0$ there exists two local minima at $(h,s)=(v,0)$ and $(0,w)$, with $v=\sqrt{-\mu_h^2/\lambda_h}$ and $w=\sqrt{-\mu_s^2/\lambda_s}$, and the former is the true vacuum, preserving the $\Z_2$ symmetry. The two minima are separated by a tree-level barrier induced by the $\lambda_{hs}$ term. Since $s$ does not acquire a vacuum expectation value at zero-temperature, it does not mix with $h$, and the physical masses of the scalars can be easily obtained as,
\be\label{tree-level_2}
m_h^2=-2\mu_h^2,\quad m_s^2=\mu_s^2+\lambda_{hs}v^2.
\ee
Given $m_h=125$ GeV and $v=246$ GeV, we only have three free parameters, which can be adopted as $\{m_s,\lambda_{hs},\lambda_s\}$. Note that $m_s>m_h/2$ is required by phenomenology to evade the $h\to ss$ bound from collider experiments.

\begin{figure}
\begin{center}
\includegraphics[width=0.99\textwidth]{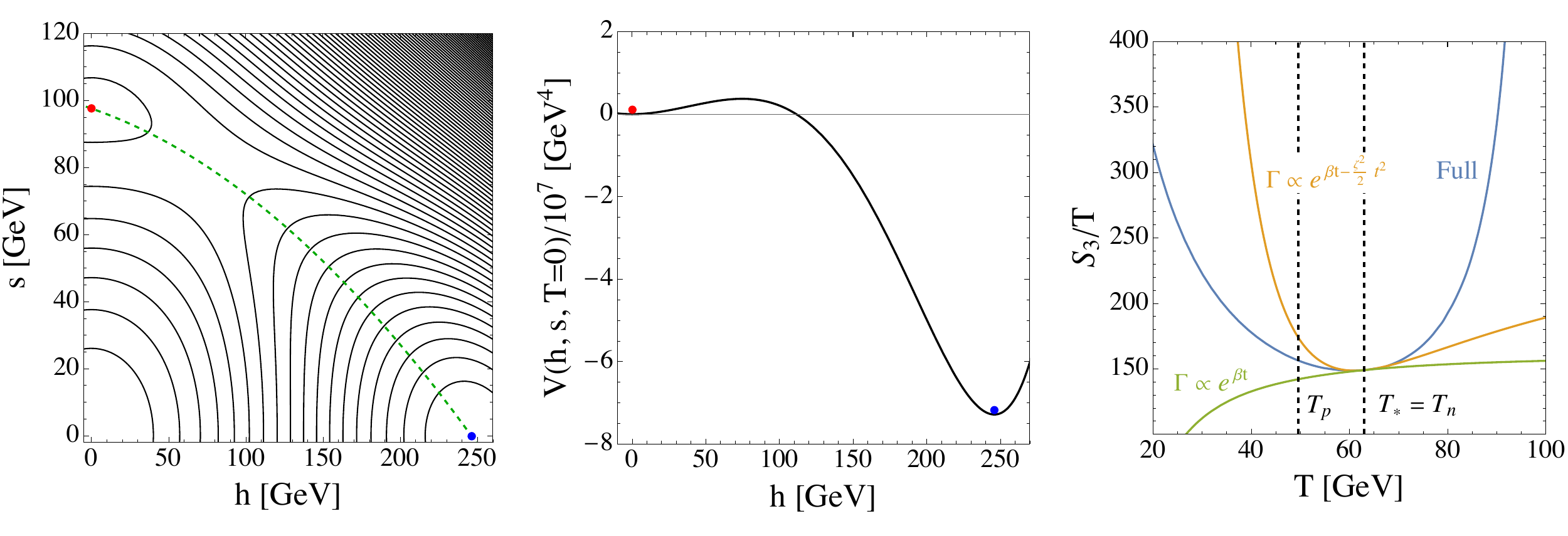}
\caption{Left: the zero-temperature potential contours $V(h,s,0)$, with the blue and red points denoting the true and false vacua, respectively. Middle: the field value along the green dashed line of the left panel. Right: the $S_3(T)/T$ curves for different treatments. The parameters are adopted as $m_s=218$ GeV, $\lambda_s=2$, and $\lambda_{hs}=1.108$.}
\label{fig:xSM_S3T}
\end{center}
\end{figure}

It is well-known that when $T>0$ the $\Z_2$-symmetric xSM can realize a two-step phase transition via the trajectory $(0,0)\to (0,w_*)\to(v_*,0)$, in which the second step is a first-order electroweak phase transition. Under the thermal potential \Eq{VT_sim}, this is the only possible FOPT pattern, and its necessary condition is~\cite{Bian:2019kmg} 
\be\label{degenerate_V}
\frac{c_s}{c_h}<\frac{\mu_s^2}{\mu_h^2}<\frac{\sqrt{\lambda_s}}{\sqrt{\lambda_h}}<\frac{\lambda_{hs}}{\lambda_h},
\ee
derived from the existence condition of $T_c$. Within this region, we resolve the FOPT dynamics of the model with the assistance of the {\tt Python} package {\tt cosmoTransitions}~\cite{Wainwright:2011kj}, which helps to evaluate $S_3(T)/T$. In Fig.~\ref{fig:xSM_S3T}, the left panel displays the contours of the potential $V(h,s,T)$ at $T=0$, with fixed values of $m_s=218$ GeV, $\lambda_s=2$, and $\lambda_{hs}=1.108$. Two distinct local minima, located at $(0,w)$ (red point, false vacuum) and $(v,0)$ (blue point, true vacuum), are separated by a barrier resulting from the $\lambda_{hs}$ term. This barrier is clearly illustrated in the middle panel, where the field values are taken along the green dashed line in the left panel. The barrier gives rise to a U-shaped $S_3(T)/T$, as depicted by the blue line in the right panel of Fig.~\ref{fig:xSM_S3T}. Additionally, we show the corresponding linear and quadratic approximations of $S_3(T)/T$. We again see the linear expansion fails to capture the shape of the complete expression.

\begin{figure}
\begin{center}
\includegraphics[width=0.63\textwidth]{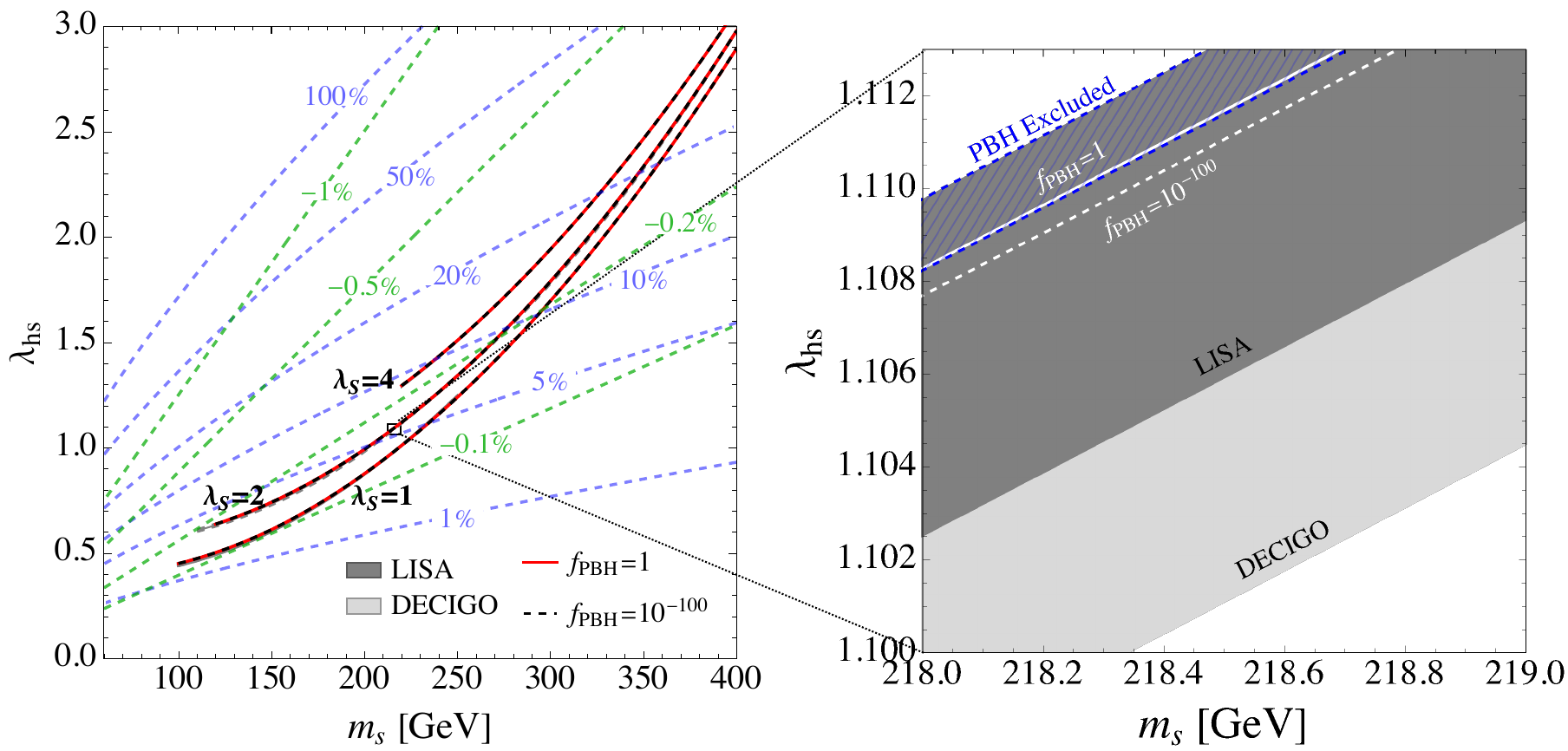}\\
\includegraphics[width=0.34\textwidth]{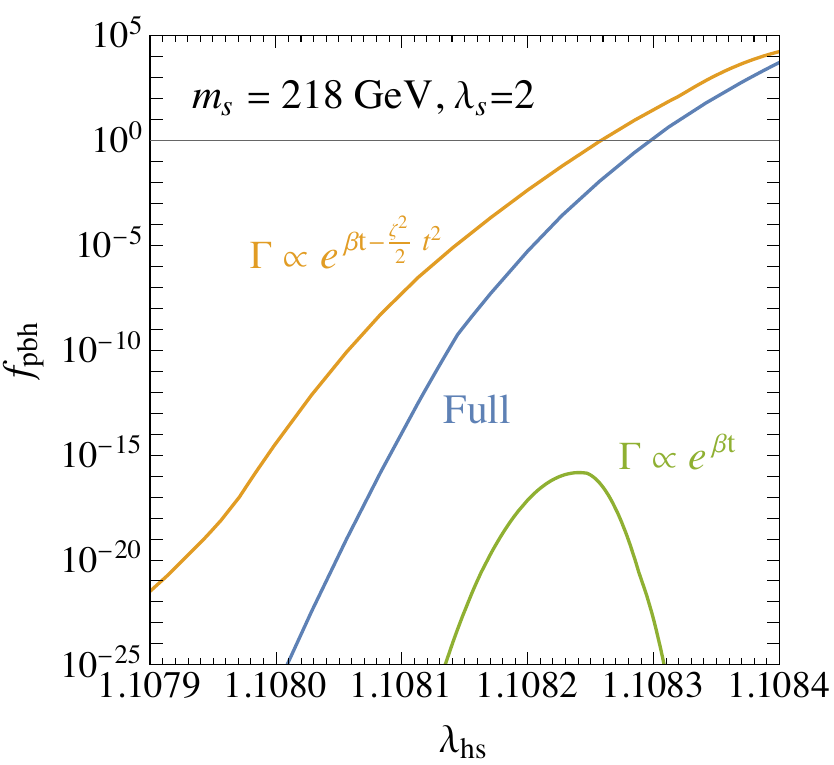}
\includegraphics[width=0.33\textwidth]{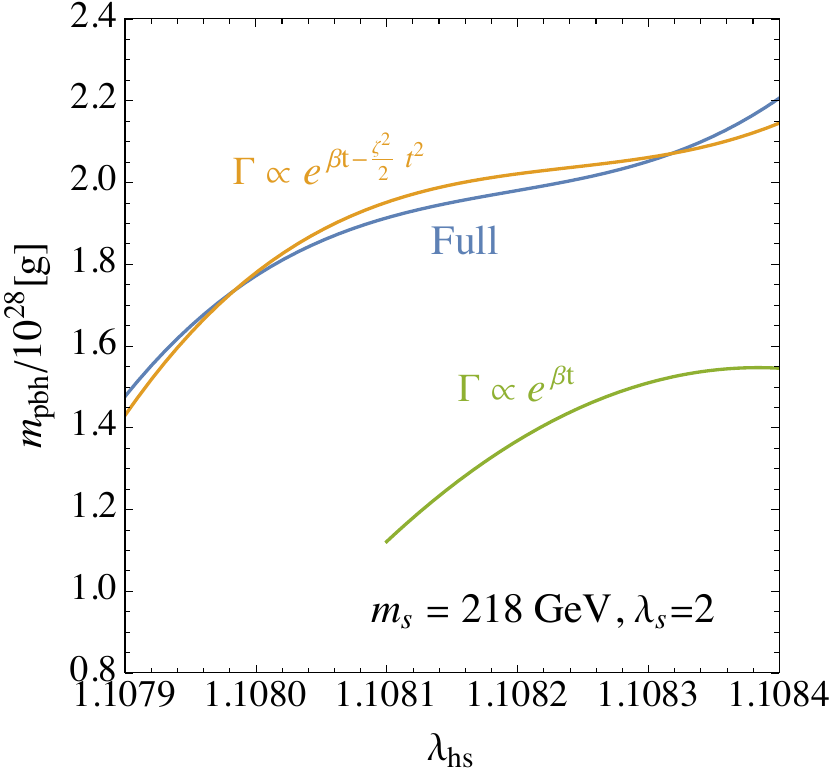}
\caption{Top: the parameter space for PBH formation in the $\Z_2$-symmetric xSM. Bottom-left (-right): The PBH fraction (mass) as a function of $\lambda_{hs}$ for different treatments of $S(t)$, where $m_s=218$ GeV and $\lambda_s=2$ are fixed.}
\label{fig:xSM_2d}
\end{center}
\end{figure}

To provide a comprehensive overview of the parameter space, we conducted a 2-dimensional scan for $m_s$ and $\lambda_{hs}$ while keeping $\lambda_s$ as different fixed values, as depicted in the top panel of Fig.~\ref{fig:xSM_2d}. In the top-left panel, we specifically plotted the parameter space for $f_{\rm pbh}\in[10^{-100},1]$ and observed that it consists of narrow regions resembling ``lines''. These ``lines'' have endpoints determined by ensuring the triviality bound~\cite{Lindner:1985uk}. It is well known that large scalar coupling constants are in general required to realize the FOPT. However, such large scalar coupling constants are strongly constrained by the triviality bound~\cite{Kanemura:2012hr}. We take into account the constraint from the triviality bound by employing mass dependent beta functions~\cite{Kanemura:2023wap}. 
We require that the Landau pole of the scalar couplings remains above 10 TeV to satisfy the triviality bound.

The top-right panel of Fig.~\ref{fig:xSM_2d} shows the details of the parameter space for $\lambda_s=2$, near the vicinity of $m_s=218.4$ GeV and $\lambda_{hs}=1.106$. The PBH formation parameter space corresponds to strong GW production, and we have evaluated the corresponding GW spectra and verified that they are easily detected by the future LISA~\cite{LISA:2017pwj}, TianQin~\cite{TianQin:2015yph}, Taiji~\cite{Hu:2017mde}, and DECIGO~\cite{Kawamura:2011zz} interferometers. Moreover, we illustrate the deviations in the $hhh$ and $hWW/hZZ$ couplings as purple and green dashed contours, respectively. These deviations can be precisely measured at future collider experiments like HL-LHC~\cite{Cepeda:2019klc}, CEPC~\cite{An:2018dwb}, ILC~\cite{Bambade:2019fyw}, and FCC-ee~\cite{FCC:2018evy}. Those future collider experiments can offer complementary and correlated probes to this PBH formation mechanism.

In the bottom-left and -right panels of Fig.~\ref{fig:xSM_2d}, we present the variations of $f_{\rm pbh}$ and $m_{\rm pbh}$ with respect to $\lambda_{hs}$, considering different treatments for $S(t)$ while keeping $m_s=218$ GeV and $\lambda_s=2$ fixed. We observe that the PBH fraction exhibits rapid changes as a function of $\lambda_{hs}$, primarily due to the significant influence of the barrier height on the probability of delayed vacuum decay. Conversely, the PBH mass $\sim10^{28} \, {\rm g} \, (\sim 10^{-5} M_{\odot})$ changes gradually as it is predominantly determined by the temperature of the FOPT. It can be seen in the figure that the linear expansion fails to accurately describe the dynamics of PBH formation, as expected.

\section{Conclusion}\label{sec:conclusion}

We explore the PBH formation through the delayed vacuum decay in the slow FOPTs, also known as the ``late-blooming'' mechanism, from a model-building perspective. We for the first time investigate this mechanism in models with zero-temperature potential barriers, taking the polynomial potential and the $\Z_2$-symmetric xSM as two examples. Our findings reveal that a U-shaped Euclidean action significantly prolongs the phase transition, leading to abundant PBH formation with sufficiently high barriers. Unlike models with classically conformal invariance~\cite{Gouttenoire:2023pxh,Salvio:2023blb,Conaci:2024tlc}, the FOPT in these models is moderate with an $\mO(1)$ $\alpha$ parameter. Therefore, our study demonstrates a general type of particle models that can realize the PBH formation without requiring ultra-supercooling. The PBH abundance is highly sensitive to model parameters, which is a distinctive feature of this mechanism and has been demonstrated in previous literature~\cite{Lewicki:2023ioy,Gouttenoire:2023naa,Gouttenoire:2023pxh,Salvio:2023blb}. Additionally, we explicitly show that the exponential nucleation approximation does not work in our models due to the U-shaped action, emphasizing the necessity for the full expression of the decay rate in particle-based studies.

Although our focus in this study is on particle models with zero-temperature barriers, the discussions presented in Section~\ref{sec:algorithm}, especially in Section~\ref{subsec:expansions} regarding the validity of the exponential approximation, also have implications for more general models, including the classically conformal ones. In those models, where $S_3(T)/T\sim T^r$ with $r>1$, satisfying \Eq{reduced_linear_con} becomes challenging if one desires a small $\beta/H_*$ to facilitate PBH formation. Consequently, the exponential approximation may also fail in PBH studies of such models, resulting in significant errors of several orders of magnitude. It is also important to note that for FOPTs characterized by long-lasting or ultra-supercooling regimes, it is necessary to verify the fulfillment of \Eq{didt} to ensure the completion of the phase transition, which is not realized in some relevant literature.

Our work can be extended in several ways. While we have chosen $v_w=1$ as a benchmark value, it is important to consider the friction force induced by plasma particles, which typically leads to bubble wall velocities smaller than 1~\cite{Lewicki:2021pgr,Laurent:2022jrs,Wang:2022txy,Wang:2023kux,Ai:2023see,Ai:2024shx}. By varying $v_w$ as different constants in our calculations, we find that lower values of $v_w$ result in later percolation and larger PBH abundance, with differences spanning several orders of magnitude. Therefore, accurately determining the velocity within a specific model is crucial for deriving the PBH abundance. Different definitions of overdensity $\delta(t)$ exist in the literature. In our research, we adopt the definition given by \Eq{delta}, which leads to a decrease in overdensity after reaching its maximum value $\delta_{\rm max}$. This is consistent with Refs.~\cite{Gouttenoire:2023naa, He:2022amv}. However, an alternative definition of $\delta(t)$, as used in Ref.~\cite{Liu:2021svg}, is given by
\begin{equation*}
\delta(t)=\frac{\rho_{\rm in}(t)}{\rho_{\rm out}(t)}\frac{a_{\rm in}^4(t)}{a_{\rm out}^4(t)}-1,
\end{equation*}
which tends to be constant after reaching $\delta_{\rm max}$. It would be interesting to further explore and compare these two definitions from a theoretical perspective. Furthermore, the finite temperature potential expression can be improved by including the full one-loop thermal corrections and beyond, and the calculation framework can be enhanced to incorporate more detailed considerations of PBH formation possibilities~\cite{Kawana:2022olo,Lewicki:2023ioy}, or by conducting numerical simulations of FOPTs~\cite{Di:2020kbw,Zhao:2022cnn}. Given the sensitivity of PBH formation to FOPT features in this mechanism, a more comprehensive treatment is expected to significantly impact the PBH abundance and the viable parameter space of a given model. However, the key qualitative conclusions drawn from our research will remain unchanged, including the relationship between zero-temperature barriers and PBH formation, as well as the breakdown of the exponential approximation.

\acknowledgments

We would like to thank Ligong Bian, Jing Liu, and Shao-Jiang Wang for the very useful and inspiring discussions. K.-P. X. also thanks the hospitality of Osaka University where part of this work was performed, and Guang-Ze Fu for the great help on programming. The work of S. K. was supported in part by the JSPS KAKENHI Grant No. 20H00160 and No. 23K17691.
M. T. was supported by the Iwanami Fujukai Foundation. K.-P. X. is supported by the National Science Foundation of China under Grant No. 12305108.

\appendix
\section{FOPT dynamics under the exponential approximation}\label{app:linear_expansion}

In this appendix, we derive the analytical results of the FOPT dynamics under the exponential approximation $\Gamma(t)\approx \Gamma_*e^{\beta (t-t_*)}$. Assume the FOPT is fast that $a(t)$ almost does not change during the transition, then the false vacuum fraction can be explicitly integrated out as
\be\label{Ft_linear}
F(t)\approx\exp\left\{-\frac{4\pi}{3}\int_{t_c}^{t}\d t'\Gamma(t')\left(\int_{t'}^t\d t''v_w\right)^3\right\}
\approx\exp\left\{-\frac{8\pi v_w^3}{3\beta^4/\Gamma_*}e^{\beta(t - t_*)}\right\},
\ee
where we have assumed $\beta(t-t_c)\gg1$ and dropped the terms proportional to $e^{-\beta(t-t_c)}$ inside the exponent. Then the bubble density can be also explicitly derived as
\be
n_b(t)\approx\int_{t_c}^{t}\d t'F(t')\Gamma(t')\approx\left(\frac{\beta}{(8\pi)^{1/3}v_w}\right)^3\left(1-F(t)\right).
\ee
Therefore, in the $t\to\infty$ limit we obtain $\bar R/[(8\pi)^{1/3}v_w]\to\beta^{-1}$, and hence $(8\pi)^{1/3}v_w/(H_*\bar R)$ approaches $\beta/H_*$~\cite{Enqvist:1991xw}.

Now we would like to derive the (rough) relation between FOPT duration and $\beta^{-1}$. Rewriting \Eq{Ft_linear} as
\be
t - t_*=\beta^{-1}\log\left[-\frac{3}{8\pi v_w^3} \frac{\beta^4}{\Gamma_*}\log F(t)\right],
\ee
and assuming the false vacuum volume fractions at $t_2$ and $t_1$ are $F_2$ and $F_1$, respectively, we find
\be
\Delta t\equiv t_2-t_1=\beta^{-1}\log\left(\frac{\log F_2}{\log F_1}\right).
\ee
An appropriate definition of the FOPT duration is to set
\be
F_2=\eta,\quad F_1=1-\eta,
\ee
where $\eta$ is a small positive number. This definition of $\Delta t$ depends logarithmically on the parameter $\eta$. For $\eta=10^{-2}$, $10^{-3}$, and $10^{-4}$, we obtain $\Delta t=6.1\,\beta^{-1}$, $8.8\,\beta^{-1}$, and $11.4\,\beta^{-1}$, respectively. Therefore, typically the FOPT duration $\Delta t\approx\mO(10)\times\beta^{-1}$, consistent with the usual interpretation of $\beta^{-1}$ in the exponential approximation.

\bibliographystyle{JHEP-2-2.bst}
\bibliography{references}

\end{document}